\documentclass{pasj00}

\begin{document}
\SetRunningHead{Iwata et al.}{CO Observation of SDSS 1044$-$0125}
\Received{0000/00/00}
\Accepted{0000/00/00}

\title{CO($J=6-5$) Observations of the Quasar SDSS1044$-$0125 at $z$ =
5.8}

\author{Ikuru \textsc{Iwata} and Kouji \textsc{Ohta}}%
\affil{Department of Astronomy, Faculty of Science, Kyoto University,
Sakyo-ku, Kyoto 606-8502} 
\email{iwata@kusastro.kyoto-u.ac.jp}

\author{Kouichiro \textsc{Nakanishi} and Kotaro \textsc{Kohno}}
\affil{Nobeyama Radio Observatory, National Astronomical Observatory of
Japan,\\
Minamimaki, Minamisaku, Nagano, 384-1305}

\and

\author{Richard G. \textsc{McMahon}}
\affil{Institute of Astronomy, Madingley Road, Cambridge,
CB3 0HA, UK}

\KeyWords{CO emission line --- Galaxies: formation --- 
Galaxies: individual(SDSSp104433.04$-$012502.2) --- 
Quasars: formation} 

\maketitle

\begin{abstract}
We present a result of the quasar CO($J=6-5$) observations of 
SDSSp J104433.04$-$012502.2 at $z$ = 5.8.
Ten-days observations with the Nobeyama Millimeter Array yielded an
rms noise level of $\sim$ 2.1 mJy beam$^{-1}$ in a frequency range 
from 101.28 GHz to 101.99 GHz at a velocity resolution of
120 km s$^{-1}$.
No significant clear emission line was detected
in the observed field and frequency range. 
Three sigma upper limit on the CO($J=6-5$) 
luminosity of the object is 
2.8 $\times 10^{10}$ K km s$^{-1}$ pc$^2$, 
corresponding to a molecular gas mass of $1.2 \times 10^{11} M_{\odot}$, 
if a conversion
factor of 4.5 $M_{\odot}$ (K km s$^{-1}$ pc$^2$)$^{-1}$ is
adopted.
The obtained upper limit on CO luminosity is slightly smaller 
than those observed in quasars at
$z=4-5$ toward which CO emissions are detected.
\end{abstract}

\section{Introduction}

A luminous quasar in a high redshift universe has been
supposed to inhabit a massive host galaxy in its forming phase.
\citet{Haehnelt(1993)} showed that this scenario
can reproduce an observed evolution of luminosity function of
AGNs/QSOs well.
More recently, \citet{Kauffmann(2000)} 
showed the star formation history of the universe can also be
accounted by this scenario with some assumptions for physical 
process of star formation and AGN formation.
The scenario is also supported by the linear correlation 
between a black hole mass at the center of a galaxy 
and a mass of a spheroid component of its host galaxy in the 
local universe (e.g., \cite{Kormendy(1995)}).
Thus a luminous quasar at high redshift may be a site where
burst star formation under which bulk of stellar population
of a present-day massive elliptical galaxy formed is on-going.

In these several years, huge amount of
molecular gas and dust have been found in quasars up to $z=4.7$ 
(e.g., \cite{Ohta(1996)}; \cite{Omont(1996a)}a;
\cite{Isaak(1994)}; \cite{Omont(1996b)}b;
\cite{McMahon(1999)}).
Molecular gases have been also detected toward
two radio galaxies at $z=3.8$ (\cite{Papadopoulos(2000)}).
These detections strongly suggest the presence of intense starburst
in the host galaxies, and the inferred star
formation rates in these systems are an order of 1000 $M_{\odot}$
yr$^{-1}$, judging from the estimated masses of the molecular gas
and the dust ($10^{10-11} M_{\odot}$ and $10^{9}
M_{\odot}$, respectively).
Physical conditions of the molecular clouds 
can also be examined by multi-transition CO observations 
(e.g., \cite{Ohta(1998)}; \cite{Papadopoulos(2001)}).
Moreover, some of these objects show the presence of two
molecular gas components physically close each other
(\cite{Omont(1996a)}a; \cite{Papadopoulos(2000)});
we may be witnessing a merging process and  the star burst may
be triggered by the interaction.
Therefore, a detection of a CO emission from a higher redshift quasar
provides us an opportunity to study starting epoch of galaxy
formation, star formation properties of a galaxy in a forming phase,
and a dynamical assembling process of a galaxy.
It may be also possible to obtain a clue to examine whether a 
quasar appeared before formation of bulk of stars in a galaxy by 
CO observations as well as submillimeter observations of quasars
at the highest redshift.

SDSSp J104433.04$-$012502.2 (or SDSS 1044$-$0125 for short) is a
quasar located at $z=5.8$, which was the highest among quasars
known at the time of the observation (\cite{Fan(2000)}).
This quasar is one of the most promising targets for CO
search, because
(1) the continuum emission at 850$\mu$m has been recently
detected with a flux density of $6.2 \pm 2.0$ mJy by SCUBA on JCMT
(McMahon, private communication).
(2) A rest-frame $B$-band absolute magnitude is estimated to be $\sim -28$ mag,
which is the brightest among high-$z$ quasars and is comparable
to those toward which CO emission was detected (\cite{Omont(1996b)}b).
(3) The rest-frame UV spectrum shows a broad "weak"-type profile of Ly$\alpha$;
CO and dust are tend to be detected in such quasars (\cite{Omont(1996b)}b).
(4) A Lyman limit system at $z=5.72$ is seen in the spectrum of the
quasar;
the presence of the Lyman limit system at such a close redshift
is also a common feature seen in high-$z$
quasars toward which CO emission is detected.
Motivated by these, we made CO($J=6-5$) observations of the
quasar aimed at pushing the CO detection back to the earlier
universe.
Throughout this paper, we adopt a cosmological parameter set of
$H_0 = 50$ km s$^{-1}$ Mpc$^{-1}$, $q_0 = 0.5$, and $\Lambda =0$.

\section{Observations and Data Reduction}

The observations were executed in total 10 nights, 
25th to 29th November 2000, and 20th to 24th December 2000, 
using Nobeyama Millimeter Array (NMA) of Nobeyama Radio Observatory
(NRO)\footnote{NRO is a branch of the National Astronomical Observatory,
an inter-university research institute operated by the Ministry of
Education, Culture, Sports, Science and Technology}. 
The array configuration for both observations was D, which is 
the most compact configuration of NMA, giving an angular
resolution (half-power beam size) of 7$^{\prime\prime}$ at the
observed frequency.
The field of view was $\sim$70$''$ centered on the quasar.

Fan et al. (2000) estimated the redshift of SDSS 1044$-$0125 
as 5.80 $\pm$ 0.02, using observed wavelengths 
of O $\rm \sc I$ + Si $\rm \sc II$ $\lambda$1302 
and Si $\rm \sc IV$ + O $\rm \sc IV$] $\lambda$1400
emission lines. It is known that high-ionization rest-frame UV
emission lines such as Si $\rm \sc IV$ observed in quasars at $z > 1$
show blueshifts of several hundred km s$^{-1}$ or larger
with respect to systemic redshifts (\cite{McIntosh(1999)}).
We adopted 500 km s$^{-1}$ for the amount of the blueshift and obtained 
an estimated systemic redshift of 5.811. 
The difference between the observed frequencies of CO($J=6-5$) emission 
from objects at the redshifts of 5.800 and 5.811 are $\sim$ 165
MHz, which is very much small as compared with a wide coverage of 
the spectrometer used (1024 MHz, $\sim 3000$ km s$^{-1}$).
The central frequency for the first observing run (November) was 
101.52 GHz, the expected frequency of CO($J=6-5$) emission 
from the object located at $z = 5.811$.
Among the data taken in the first observing run, some data showed a
hint of the presence of an emission feature around 101.8 GHz,
hence we slightly shifted the observed frequency for the second
observing run (December).
The central frequency was set to be 101.76 GHz.
The frequency range covered by both observing runs was
$\sim$780 MHz (about 2300 km s$^{-1}$ or a redshift ranging from
5.78 to 5.83). 

In general the weather conditions were 
fairly good, except for four nights, 
when the atmospheric phase stability was not always good.
The system noise temperature (SSB) was $\sim$400 K during the
observations. 
We used a quasar J1058+015 as a reference calibrator,
which is located very close ($\sim 5^\circ$) to the quasar.

Data reduction and analysis were made using 
UVPROC2 developed by NRO,
and AIPS developed by NRAO, USA.
Data reduction was made independently by two of the authors
(II and KN), by employing a different selection criterion for
quality of the data.
K.N. severely judged the quality of the data and discarded 
low quality data such as those taken under a relatively bad 
atmospheric condition or at low elevation.
I.I. adopted a relatively loose criterion for data quality 
and included data of all 10 days, and the resulting total
effective on-source time was $\sim$ 36 hours.
There is no distinct difference between the resulting two spectra
at the quasar position, 
except for a rather poorer signal-to-noise ratio in the severe
criterion case.
We also examined the two dirty channel maps and 
found no significant difference at and around the quasar.
We will use the latter result below.

\section{Results and Discussion}

\begin{figure}
  \begin{center}
    \FigureFile(80mm,80mm){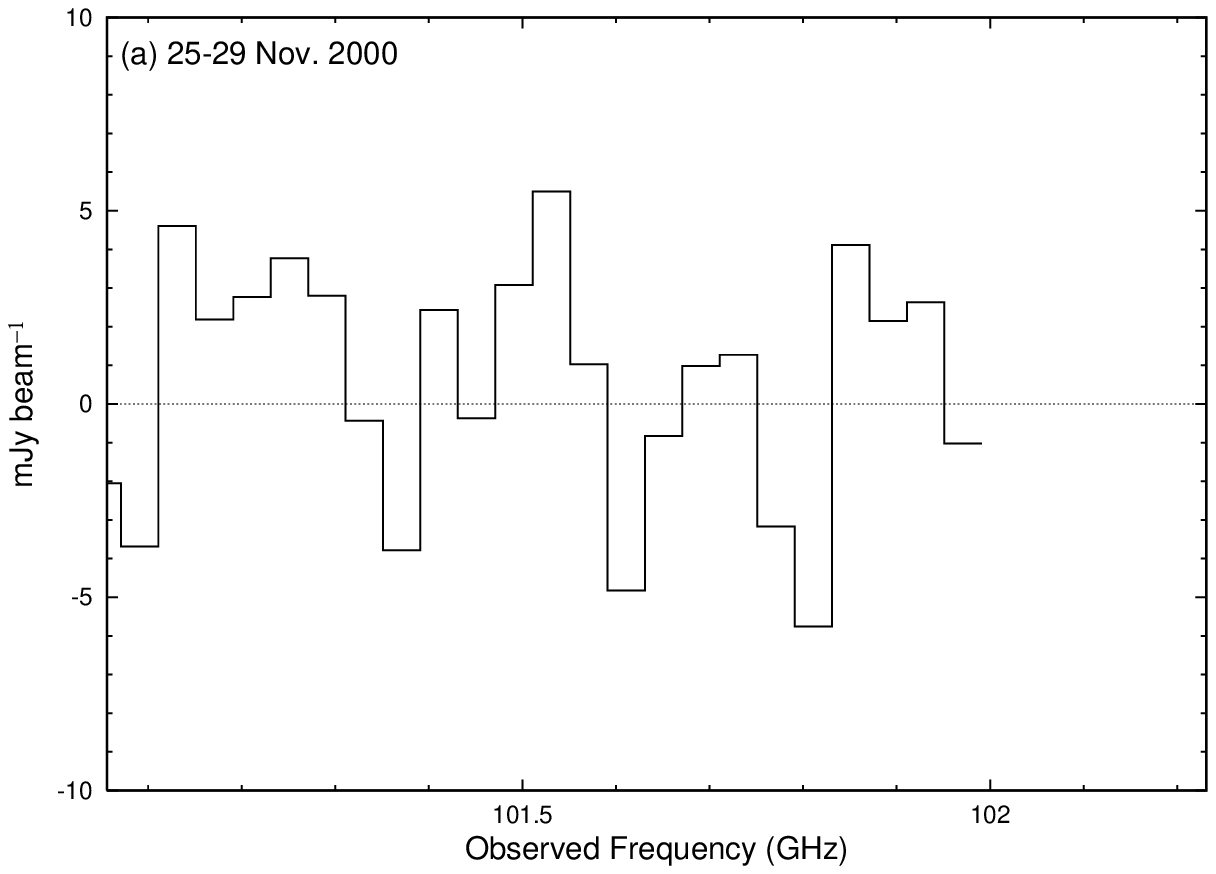}
    \FigureFile(80mm,80mm){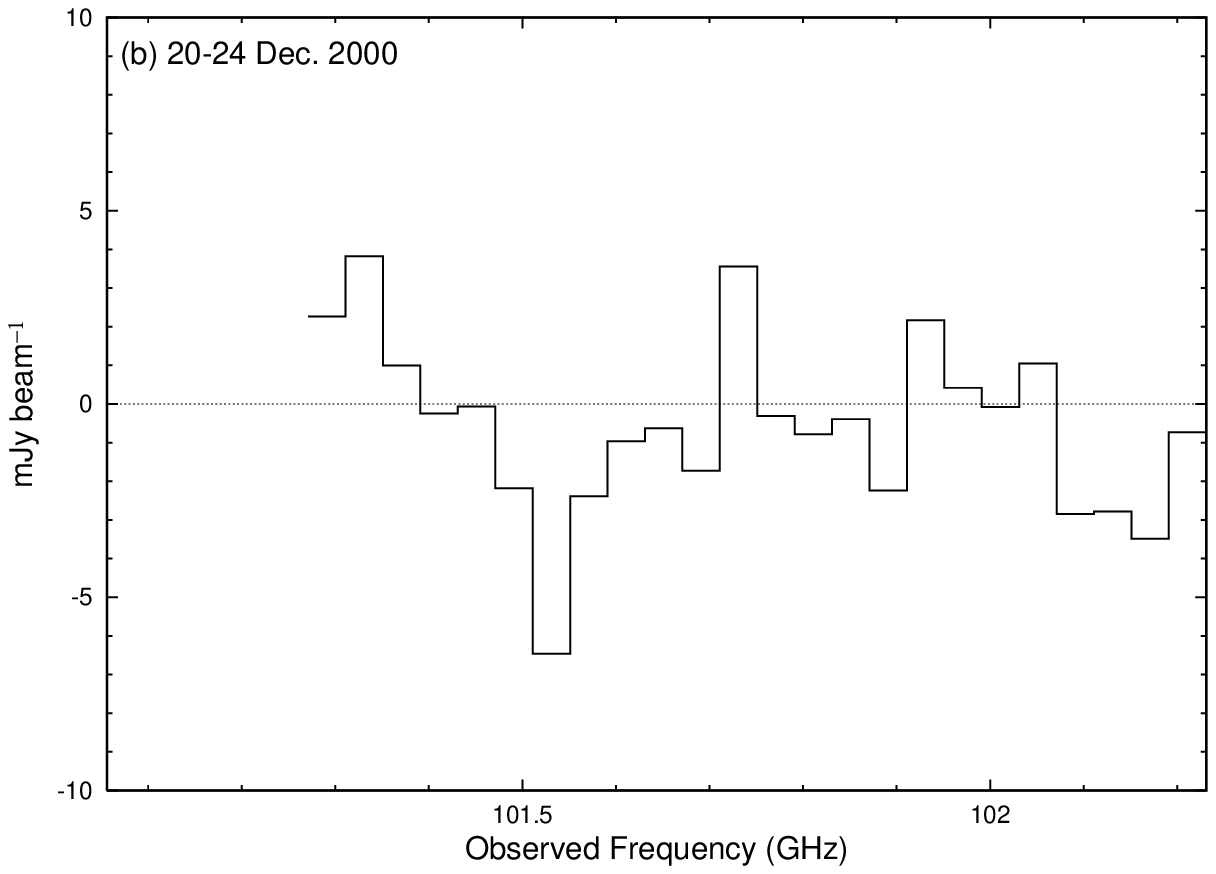}
    \FigureFile(80mm,80mm){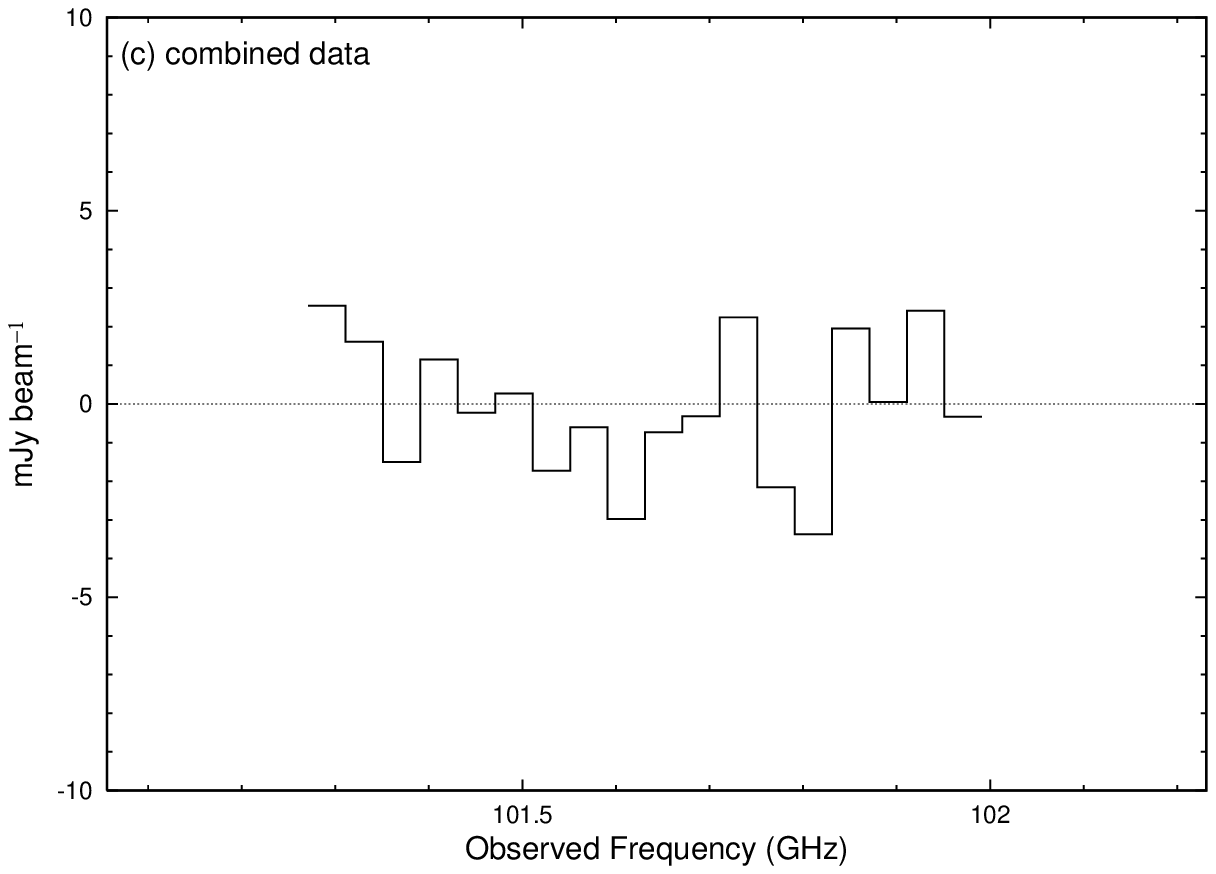}
  \end{center}
  \caption{Spectra at the optical position of the quasar
 SDSS1044$-$0125. (a) and (b) show spectra obtained
 during the first and the second observing run, respectively.
 (c) A spectrum made by combining 10 days data.
 Only the frequency range common to both observing runs 
 is plotted. 
}
  \label{fig1}
\end{figure}

In figure \ref{fig1} we present spectra at the optical position 
of the quasar
with a resolution of 40 MHz ($\sim$ 120 km s$^{-1}$).
Figure \ref{fig1}(a) and (b) show the spectra obtained in the first
observing run and in the second one, respectively. 
Note that the observed frequency range for the
two runs is slightly different from each other. 
For each spectrum, data of five-days are combined.
Resulting rms noise at a 40 MHz bin is  
3.2 mJy beam$^{-1}$ and 2.7 mJy beam$^{-1}$, respectively.
There is no significant emission feature appeared in both spectra. 
Figure  \ref{fig1}(c) shows a spectrum made by combining 10-days data.
Only the frequency range common to both observations 
is plotted, and a typical rms noise is 2.1 mJy beam$^{-1}$.
Within the reduced rms noise level, figure \ref{fig1}(c) indicates
no significant emission feature.
Although we also checked spectra with different frequency resolutions, 
we could not find any significant sign of emission lines.

We carefully examined the channel maps of each observing run and
combined data with different channel binings.
There are some ``spikes'' in the field around the quasar 
which appeared in data obtained taken in particular one or two days.
Some of them remain in combined data.
We consider they are spurious noises, since
none of such features appears throughout the other observing days.

The 3$\sigma$ upper limit on the CO luminosity ($L'_{\rm CO}$) of the
quasar is calculated by modifying the equation by 
Solomon, Downes, and Radford (1992) as follows,

%
\begin{equation}
L'_{\rm CO} = \frac{c^2}{2k} \frac{\ d_L^2 \ 3 \sigma_\nu\ 
\sqrt{\delta v \Delta v}}{ \nu_{rest}^2\  (1 + z)}\ ,
\end{equation}
where 
$c$ is the speed of light, 
$k$ the Boltzmann constant, 
$d_L$ the luminosity distance to the object,
$\sigma_\nu$ the rms noise flux density (2.1 mJy beam$^{-1}$ for a
frequency range 101.28 GHz to 101.99 GHz), 
$\delta v$ the velocity resolution (120 km s$^{-1}$),
$\Delta v$ the assumed velocity width of the emission line (250 km s$^{-1}$), 
$\nu_{rest}$ the rest-frame central frequency observed,  
and $z$ the redshift of the object.
The 3$\sigma$ upper limit of the CO($J=6-5$) luminosity 
of SDSS1044$-$0125 is thus derived to be 2.8 $\times 10^{10}$ K km s$^{-1}$ pc$^2$.
If we adopt a CO-to-H$_2$ conversion factor 
of 4.5 $M_{\odot}$ (K km s$^{-1}$ pc$^2)^{-1}$ (e.g., Solomon et al. 1992),
the upper limit on the molecular gas mass is estimated to be $1.2 \times
10^{11}$ $M_{\odot}$.
The CO luminosity upperlimits for the frequency ranges covered by 
just one observing run are 4.2 $\times 10^{10}$ K km s$^{-1}$ pc$^2$ 
in 101.04 GHz to 101.28 GHz, and 
3.5 $\times 10^{10}$ K km s$^{-1}$ pc$^2$ in 101.992 GHz to 102.272 GHz.

We can now compare the values with those detected in high 
redshift ($z>4$) quasars.
Since the CO($J=7-6$) and CO($J=5-$4) flux densities of
BR1202$-$0725 at $z=4.7$ are very similar to each other 
(\cite{Omont(1996a)}a), 
it would be reasonable to regard that the CO($J=6-5$) flux density of 
the object is close to that of CO($J=5-4$) emission (9.3 mJy).
Then the CO($J=6-5$) luminosity of BR1202$-$0725
is estimated to be 5.1 $\times 10^{10}$ K km s$^{-1}$ pc$^2$.
(We also suppose that the line width of 220 $\pm$ 74 km s$^{-1}$
does not change with the transitions.)
The upper limit of CO($J=6-5$) luminosity of SDSS1044$-$0125 is slightly 
smaller than that of BR1202$-$0725.
CO($J=5-4$) emission from BRI1335$-$0417 at $z=4.4$ (\cite{Guilloteau(1997)})
and from BRI0952$-$0115 at $z=4.4$ (\cite{Guilloteau(1999)}) are detected.
As is the case for BR1202$-$0725, if we suppose the CO($J=6-5$) flux
densities of them are close to the CO($J=5-4$) flux densities 
(peak flux of 6mJy, velocity width of 420 $\pm$ 60 km s$^{-1}$ 
for BRI1335$-$0417, 
4mJy, 230 $\pm$ 30 km s$^{-1}$ for BRI0952$-$0115), 
then the CO($J=6-5$) luminosities for BRI1335$-$0417 and BRI0952$-$0115
are $4.3 \times 10^{10}$ and $1.6 \times 10^{10}$
K km s$^{-1}$ pc$^2$, respectively.
SDSS1044$-$0125 is less luminous than BRI1335$-$0417, 
though our 3$\sigma$ upper limit is slightly larger than 
the CO luminosity of BRI0952$-$0115.
Note, however, that there is a possibility of amplification by gravitional
lensing for these quasars, especially for BRI0952$-$0115.
Much deeper CO observations would be required to discuss the amount of
the molecular gas in this object.

\end{document}